\documentclass[review]{elsarticle}
  \usepackage{mathastext}
\usepackage{mathtools}
\usepackage{textcomp}
\usepackage{tikz}
\usepackage{soul}
\usepackage{subcaption}
\usepackage{hhline}
\usepackage{multirow}
\usepackage{tcolorbox}
\usepackage[export]{adjustbox}
\usepackage{graphicx}
\usepackage{bm}
\usepackage{float}
\usepackage{graphicx}
\usepackage{subcaption} 
\usepackage{caption}
\usepackage{subcaption}
\usepackage{rotating}
\usepackage{array,multirow}
\usepackage{lineno,hyperref}
\usepackage{color}
\usepackage{psfrag}
\usepackage{afterpage}
\usepackage{tikz}
 
\hypersetup{pageanchor=false}
\usepackage[crop=off]{auto-pst-pdf}% With an onstalled Perl always crop=on!
\ifpdf
\usepackage{forest}
\else
\usepackage{pst-barcode}
\fi
\modulolinenumbers[1]
\journal{NDT \& E International}
\begin{document}
\begin{frontmatter}
%%%%%%%%%%%%%%%%%%%%%%%%%%%%%%%%%%%%%%%%%%%%%%%%%%%%%%%%%%%%%%%%%%%%%%%%%%%%%%%%%%%%%%%%%%%%%%%%%%%%%%%%%%
\title{Effect of scanning acceleration on the leakage signal in magnetic flux leakage type of non-destructive testing}
\author[1]{Lintao Zhang\corref{mycorrespondingauthor}}
\cortext[mycorrespondingauthor]{Corresponding author}
\ead[url]{L.Zhang@swansea.ac.uk}
\author[1]{Ian M. Cameron}
\author[2]{Paul D. Ledger}
\author[1]{Fawzi Belblidia}
\author[3]{Neil R. Pearson}
\author[4]{Peter Charlton}
\author[1]{Johann Sienz}
\address[1]{Advanced Sustainable Manufacturing Technologies (ASTUTE 2020) Operation, College of Engineering, Swansea University, Bay Campus, Fabian Way, Swansea SA1 8EN, UK}
\address[2]{College of Engineering, Swansea University, Bay Campus, Fabian Way, Swansea SA1 8EN, UK}
\address[3]{Eddyfi Technologies UK Ltd., Clos Llyn Cwm, Swansea Enterprise Park, Swansea SA6 8QY, UK}
\address[4]{University of Wales Trinity Saint David,  Mount Pleasant, Swansea SA1 6ED, UK}
%%%%%%%%%%%%%%%%%%%%%%%%%%%%%%%%%%%%%%%%%%%%%%%%%%%%%%%%%%%%%%%%%%%%%%%%%%%%%%%%%%%%%%%%%%%%%%%%%%%%%%%%%%
\begin{abstract}
This novel work investigates the influence of the inspection system acceleration on the leakage signal in magnetic flux leakage type of non-destructive testing. The research is addressed both through designed experiments and simulations. The results showed that the leakage signal, represented by using peak to peak value, decreases between 20\% and 30\% under acceleration. The simulation results indicated that the main reason for the decrease is due to the difference in the distortion of the magnetic field for cases with and without acceleration, which is the result of the different eddy current distributions in the specimen. The findings will help to allow the optimisation of the MFL system to ensure the main defect features can be measured accurately during the machine acceleration. It also shows the importance of conducting measurements at constant velocity, wherever possible.
\end{abstract}
\end{frontmatter}
%\linenumbers
\section{Introduction}
\label{sec:intro}
The effective non-destructive testing (NDT) is a technique that can help prevent disasters similar to the Buncefield incident \cite{2005b}. Many NDT methods have been developed. Here, we only focus on the magnetic flux leakage (MFL) method. The principle of MFL method is based on the followings: the discontinuity of the geometry (ferromagnetic material) can cause the leakage of the magnetic field and this leakage can be captured by magnetic sensors. This leakage signal is used to predict the defect features. This technique was extensively applied to examination of defects in pipelines, pressure vessels, and specific `train' wheels in the 1960s. The defect characteristics, such as shapes, dimensions and locations, can be determined by the leakage signals and a large amount of relevant numerical and experimental research has been carried out to link the signal to the defect shape. Magnetic techniques for covering large areas generate eddy currents in the conducting permeable specimens due to the velocity of the measurement system. The magnetic Reynolds number for this type of problem is in the order of hundreds ($>>$ 1), which indicates the effect of eddy current cannot be neglected \cite{paper}. The distortion of the magnetic field under different scanning velocity is widely reported \cite{2018agan, 1997yshin, 1995gk, 2014pwang, 2006yli, 2018neil} and it is sensitive to the specimen movement direction, as demonstrated in our previous research \cite{2015lz} and is independent to the orientation of the magnetizing source. This signal deformation can influence the efficiency of the NDT, especially for determining the defect location and its severity. With the aim of compensating for this leakage signal deformation, a scheme was validated against experimental results \cite{2004gspark}. In practice, there are at least three stages, in terms of machine velocity,  whilst the machine is measuring a specimen: the acceleration stage, the constant velocity scanning stage and the decelerating stage.\\
From the short literature review above, all previous research has focused on the steady scanning stage. This opens an important question: will the MFL system scanning acceleration influence the leakage signal? Clarifying this will help to optimize the system to ensure that defect features that are scanned whilst the machine is accelerating can be measured accurately. Surprisingly, there is no previously published work in this area, according to the authors' knowledge and the principal novelty of this work is to address this question. Perhaps no previous work has been published on this subject since the time of the acceleration and deceleration stages is short. In bulk storage tank inspection, many scans are conducted with many starts and stops. This is due to the relatively small plate geometries, in the region of 10's m as compared to the kilometre of scanning in piggable pipeline applications.  This paper focuses on the application of MFL on bulk storage tank floors where frequent, separate scans are conducted.\\
The present paper is organised as follows. The experimental facility and the procedures are introduced in section \ref{sec:efp}. The numerical set-up is discussed in section \ref{sec:ns}. In sections \ref{sec:bmf} and \ref{sec:by}, the background magnetic field for different scanning velocities and the overview of the background magnetic field distribution for both with and without system acceleration are discussed. In section \ref{sec:ae}, the discussion of the acceleration influence is presented. The main conclusions and future work are included in section \ref{sec:con}.
\section{Experimental facilities and numerical set-up}
\subsection{Experimental facilities and procedures}
\label{sec:efp}
\begin{figure}
\centering
\includegraphics[width=3in]{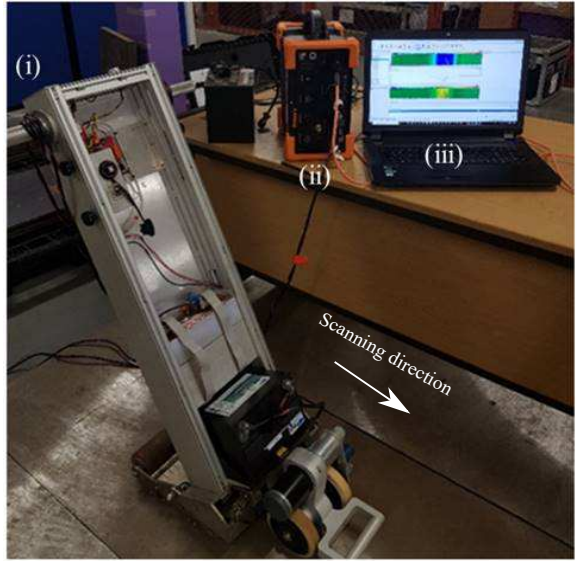}
%\vspace{-3mm}
\caption{The experimental facilities and set-up: (i) modified Floormap 3Di, (ii) Eddyfi Ectane 2 multitechnology test instrument and (iii) Eddyfi Magnifi data acquisition and analysis software.}
\label{fig:sketch}
\end{figure}

The adopted experimental facilities are identical to the work of Pullen et.al. \cite {2018neil} and their set-up is shown in Fig.\ref{fig:sketch}.
A commercial magnetic flux leakage system was adapted to conduct the experiments. The machine has a scanning velocity magnitude range between 0.5 m/s and 1 m/s. A high-resolution magnetic sensor array comprising of 64 channels is placed between the poles with fixed lift-off values relative to the specimen and the signal is transferred to the data acquisition system. A 1010 grade mild steel plate (0.5 m $\times$ 1.15 m $\times$ 6 mm) was chosen as the specimen. The reason for this selection is that 1010 grade steel is a typical parent material for storage tank floors and this thickness ensures the plate can be saturated under the current MFL assembly. Four artificial cone shape defects were manufactured with the maximum defect depths: 1.2 mm (20\% of plate thickness 6 mm), 2.4 mm (40\%), 3.6 mm (60\%) and 4.8 mm (80\%), respectively. The defects are uniformly distributed with an interval of 0.1 m. The MFL system's velocity and acceleration were determined from data captured using a position encoder.\\
Three sets of experimental trials (Trial A, B \& C) were conducted and they are summarised as follows:
\begin{enumerate}
\item Trial A: the machine was used to measure the magnetic flux leakage from a defect free specimen at three scanning velocities magnitudes (0.5 m/s, 0.75 m/s and 1 m/s). The aim for Trial A was to determine the background magnetic field for typical measurement velocities.
\item Trial B: the machine was used to measure the magnetic flux leakage from a specimen with defects at three scanning velocities magnitudes (0.5 m/s, 0.75 m/s and 1 m/s). The aim for Trial B was to determine the magnetic flux leakage for defects at constant velocity.
\item Trial C: the machine was used to measure a specimen with defects. The experiments were performed such that the system was accelerating when the sensor was passing over the defects.  The aim for Trial C was to determine the magnetic flux leakage using an accelerating MFL system. 
\end{enumerate}
For Trials B and C, the leakage was obtained from all four defect depths (1.2 mm, 2.4 mm, 3.6 mm and 4.8 mm) located as both top (near side) and bottom (far side) surface defects.
\subsection{ Numerical set-up}
\label{sec:ns}
\begin{figure}
\centering
\includegraphics[width=4in]{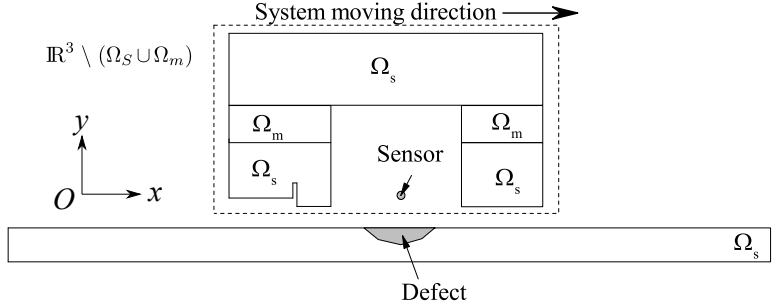}
%\vspace{-10mm}
\caption{Diagram of the experiment (not scaled): the MFL system is moving and the plate is fixed. $\Omega_m$ and $\Omega_s$ denote the domains of magnet and the steel.}
\label{fig:dia_real1}
\end{figure}

\begin{figure}
\centering
\includegraphics[width=3in]{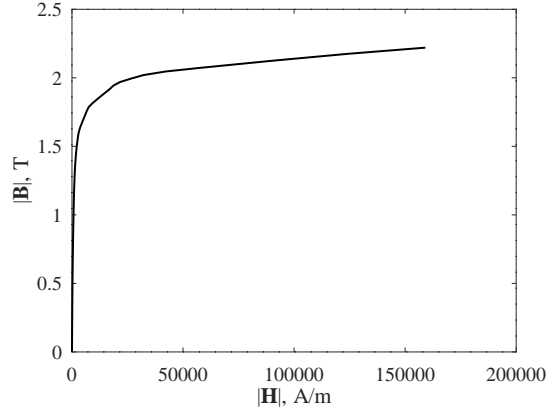}
%\vspace{-3mm}
\caption{\textbf{B}-\textbf{H} curve for steel 1010. Steel 1010 is adopted as the specimen material.}
\label{fig:dia1}
\end{figure}

Fig.\ref{fig:dia_real1} shows the diagram of experiment.
For the eddy current problem involving a conductor, the Maxwell equations simplify to:
\begin{equation}
\nabla \times \textbf{E} = - \frac{\partial \textbf{B}}{\partial t},
\end{equation}
\begin{equation}
\nabla \times \textbf{H} = \sigma \textbf{E} + \sigma (\textbf{V} \times \textbf{B}),
\end{equation}
\begin{equation}
\nabla \cdot \textbf{B} = 0,
\end{equation}
where \textbf{E}, \textbf{B}, $t$, \textbf{H}, $\sigma$, \textbf{V} are electric field, magnetic flux density, time, magnetic field, the electric conductivity and the moving system velocity, respectively. The transmission conditions are:
\begin{equation}\label{eq:t1}
[\textbf{n} \times \textbf{H}] = \textbf{0},
\end{equation}
\begin{equation}\label{eq:t2}
[\textbf{n} \times \textbf{E}] = \textbf{0},
\end{equation}
where \textbf{n} is the unit vector outward normal. The decay condition, applied on the interface between conducting and non-conducting regions, is:
\begin{equation}\label{eq:de1}
\textbf{H} = O(|\textbf{x}|^{-1}),  \textbf{ as } |\textbf{x}| \to \infty,
\end{equation}
where `[ ]' denotes the jump (e.g. between the plate and free space) and \textbf{x} is the coordinate vector. NdFe52 is selected as the magnet material and steel 1010 is selected as the bridge, poles and the plate's material. Fig.\ref{fig:dia1} shows the \textbf{B}-\textbf{H} curve for steel 1010.
It shows the non-linear constitutive behaviour, {\bf B} ={\bf B}(H) in $\Omega_s$. In $\Omega_m$, we have {\bf B}= $\mu_r \mu_0$\textbf{H} with $\mu_r$ =1.43 and in ${\rm I\!R}^3$ $\setminus$ ($\Omega_S \cup \Omega_m)$ then we have simple relationship {\bf B}= $\mu_0 {\bf H}$ where $\mu_0$ is the permeability of free space. 
\subsubsection{Two-dimensional simulation}
\label{sec:ns1}
\begin{figure}
\centering
\includegraphics[width=4in]{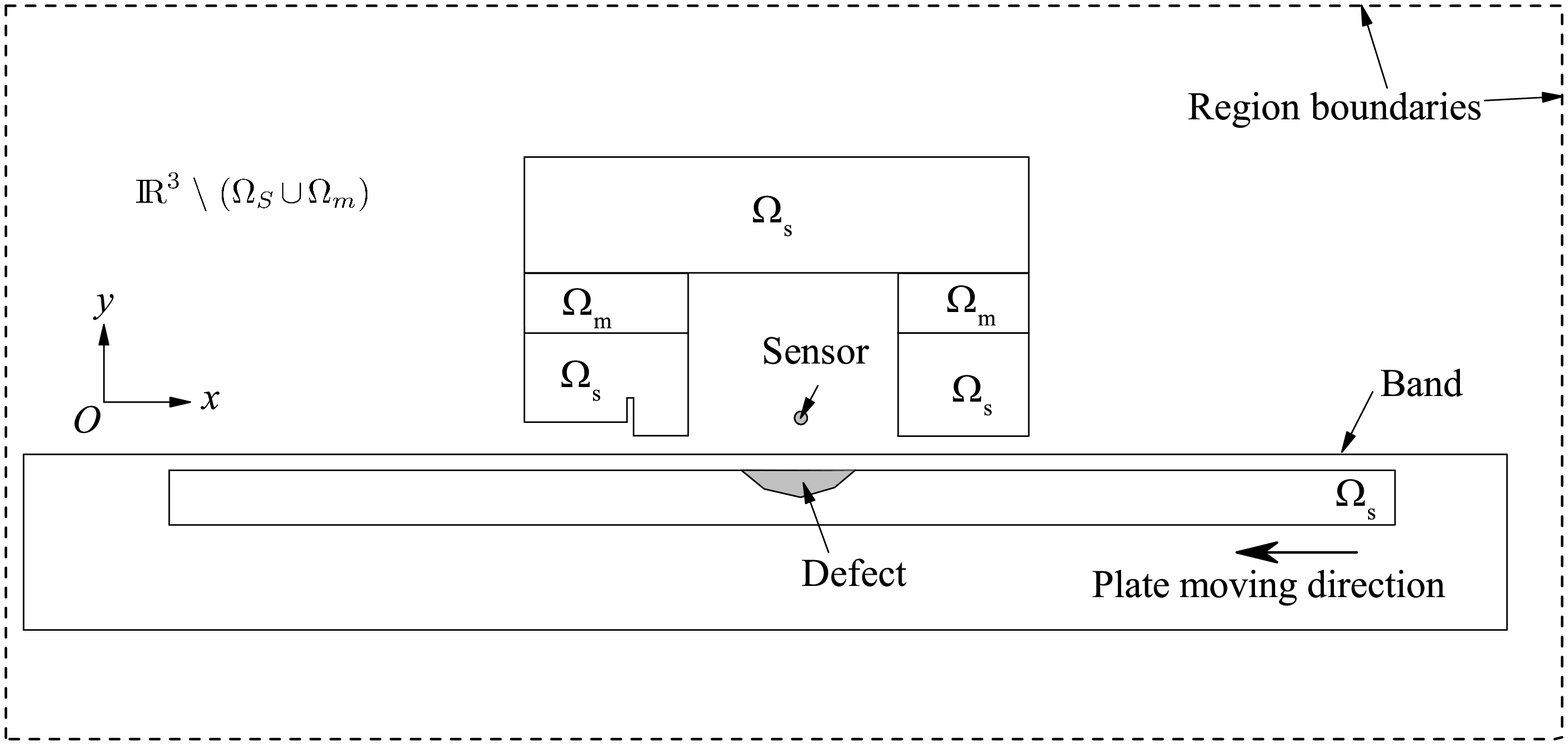}
%\vspace{-3mm}
\caption{Diagram of the problem domains for modelling: region is the simulation domain and the objects in the band area can be assigned a moving velocity.}
\label{fig:dia}
\end{figure}

For 2D problem, the control equations for the different subdomains reduce to:
\begin{equation}
\nabla \times (\textbf{H} (\nabla \times \textbf{A}))= - \sigma_s  \frac{\partial \textbf{A}} {\partial t} + \sigma_s \textbf{V} \times \nabla \times \textbf{A} \textrm{  in } \Omega_s,
\end{equation} 
\begin{equation}
\nabla \times (\mu_r\mu_0)^{-1} \nabla \times \textbf{A} = \nabla \times \textbf{H}_c \textrm{  in } \Omega_m,
\end{equation} 
\begin{equation}
\nabla \times \mu_0^{-1} \nabla \times \textbf{A} = 0 \textrm{  in } {\rm I\!R}^3 \setminus (\Omega_s \cup \Omega_m),
\end{equation}
where $\sigma_s$ and $\textbf{H}_c$ are electric conductivity for steel and the magnetic coercivity. For the current case, we have $\sigma_s$ = 2$\times$10$^6$ S/m and $|\textbf{H}_c|$= 7.96$\times$10$^5$ A/m. At the material interfaces, the transmission conditions are:
\begin{equation}
[\textbf{n} \times \textbf{A} ] =  \textbf{0} \textbf{ on } \partial \Omega_a \cap  \partial \Omega_b,
\end{equation}
\begin{equation}
[\textbf{n} \times \mu^{-1} (\nabla \times \textbf{A})] = \textbf{0} \textbf{ on } \partial \Omega_a \cap  \partial \Omega_b,
\end{equation}
where $\Omega_a$ and $\Omega_b$ represent different materials in the model. The decay condition for the 2D model is:
\begin{equation}
\textbf{A} = O(|\textbf{x}|^{-1}),  \textbf{ as } |\textbf{x}| \to \infty.
\end{equation}
Note that for 2D problems, we have \textbf{A}= $A_z(x,y) \textbf{e}_z$, where $\textbf{e}_z$ is the unit vector along $z$ direction.
Fig.\ref{fig:dia} shows the diagram of simulation.
The decay condition is approximated by the balloon boundary condition, applied on the region edges, in the simulation: the $z$ component of the magnetic vector potential, $A_z$, goes to zero at infinity. Note that in the simulations the bridge is fixed in position and the plate is moving, which is the opposite to the real situation. However, the overall effect is the same.\\
We employ ANSYS Maxwell finite element package for the approximate solution of the system described above. This includes a Newton-Raphson algorithm for dealing with the non-linear constitutive relationship in $\Omega_s$. We set the required tolerance for this iterative scheme to be such that the relative residual is smaller than 0.0001. Time integration of the transient system is achieved by Runge-Kutta scheme (third order). In the work of Zhang etc.\cite{2015lz} a mesh size sensitivity for a similar problem has already been conducted. It was established that a minimum mesh (mesh type: triangle) density of 0.5 mm (element maximum length between two poles) and a time step size of 0.0005 s is sufficient to achieve reliable results for this problem and is employed also here.
\subsubsection{Three-dimensional simulation}
The governing equations for 3D problem can be modified as follows:
\begin{equation}
\nabla \times \sigma_s^{-1}  \nabla \times \textbf{H} + \frac{\partial \textbf{B}(\bf H)}{\partial t}= \textbf{0} \textrm{  in } \Omega_s,
\end{equation}
\begin{equation}
\nabla \times \textbf{H} = \nabla \times \textbf{H}_c \textrm{  in } \Omega_m,
\end{equation}
\begin{equation}
\nabla \times \textbf{H} = \textbf{0} \textrm{  in } {\rm I\!R}^3 \setminus (\Omega_m \cup \Omega_s),
\end{equation}
\begin{equation}
\nabla \cdot \textbf{B} = 0 \textrm{  in } {\rm I\!R}^3.
\end{equation}
The transmission and decay conditions are as in Eq.\ref{eq:t1}, \ref{eq:t2} and \ref{eq:de1}. The simulation of the 3D problem is also performed using the ANSYS Maxwell solver using a similar setup to the 2D problems described in Section \ref{sec:ns1} apart from the use of physical fields rather than a vector potential formulation. Compared to the 2D balloon conditions, zero tangential \textbf{H} field is applied at the region surfaces and the domain is chosen to be sufficient large. Since the problem has a symmetry to $x-y$ plane, only half the problem is modelled to reduce computational expense. Therefore, a symmetry boundary condition (magnetic flux tangential) is applied at the middle surface of the whole domain. For this 3D complicated geometry problem, it is hard to conduct the mesh sensitivity test. A minimum mesh (mesh type: tetrahedron) density of 1 mm (element maximum length between two poles). The simulation time step is 0.001 s. Newton-Raphson algorithm for dealing with the non-linear constitutive relationship in $\Omega_s$ and the nonlinear residual is 0.005. The scalar potential shape function is second order.
\section{Results and discussion}
\subsection{Background magnetic field}
\label{sec:bmf}
\begin{figure}[htp!]
\centering
\begin{subfigure}[b]{0.49\textwidth}
\includegraphics[width=\textwidth]{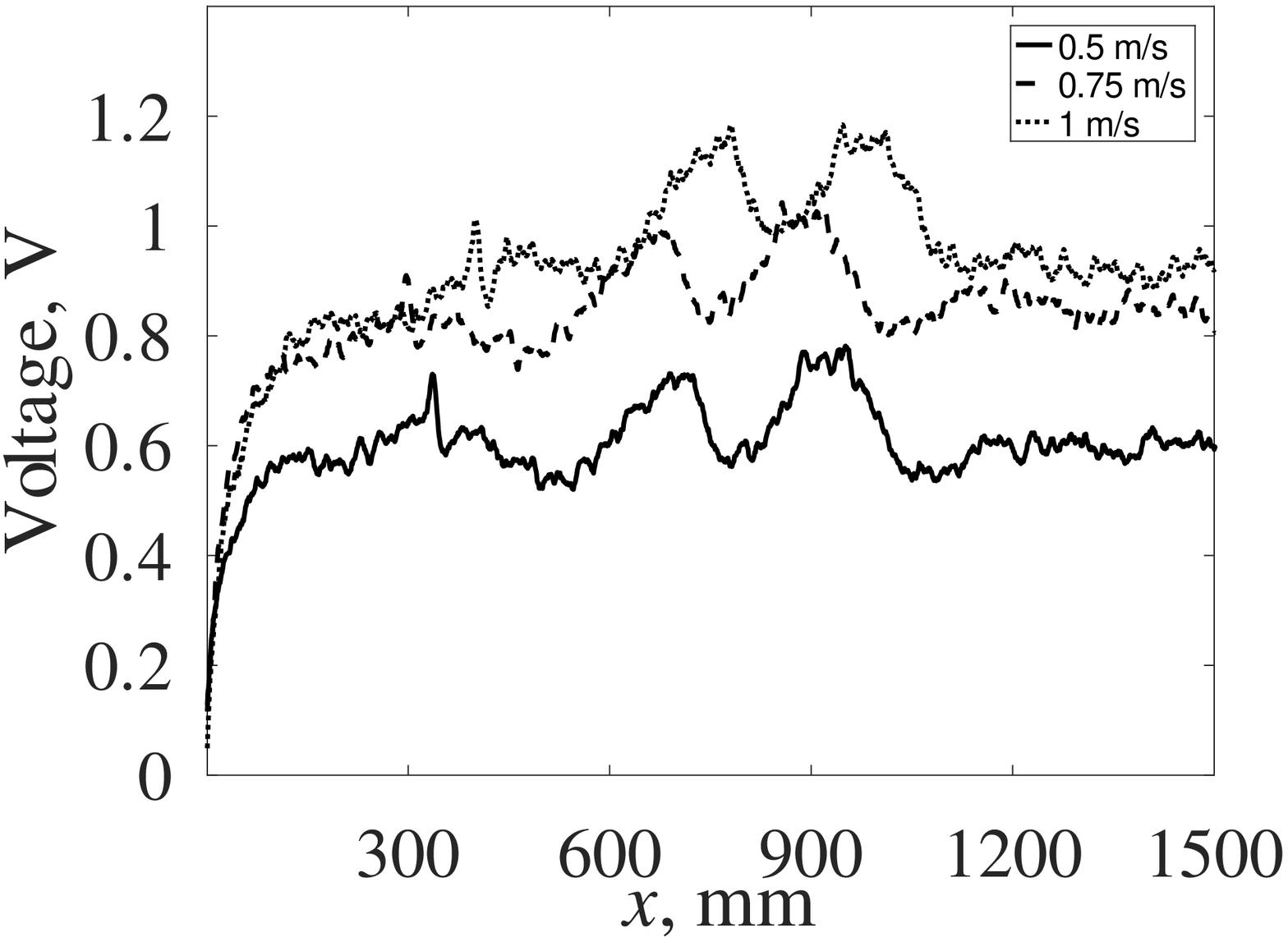}
\caption{}
\end{subfigure}
\begin{subfigure}[b]{0.49\textwidth}
\includegraphics[width=\textwidth]{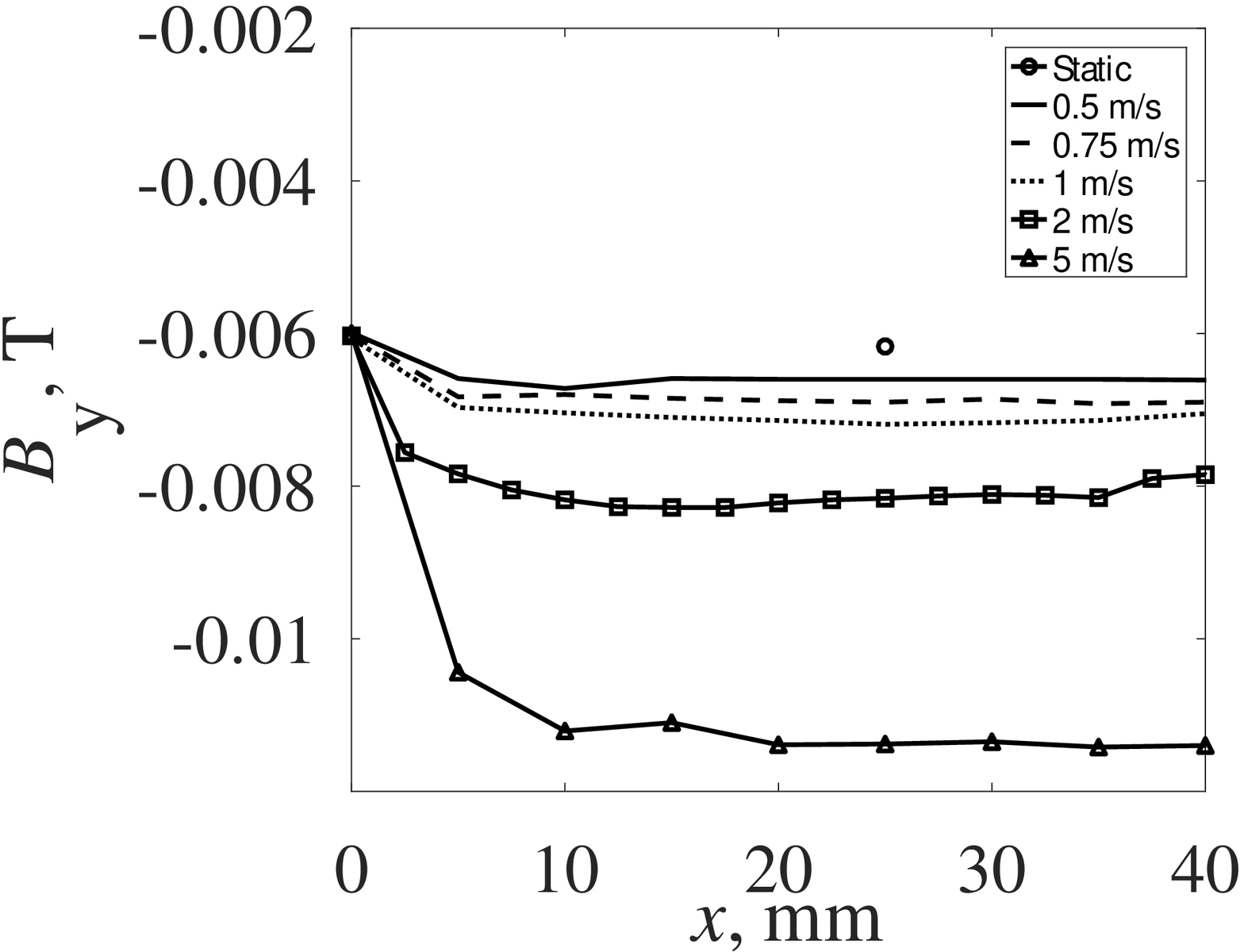}
\caption{}
\end{subfigure}
%\vspace{-3mm}
\caption{(a) Experimental results: leakage signal at different scanning velocities for the defect free specimen.  (b) Numerical results: \textit{B}$_y$ signal at different scanning velocities for the defect free specimen. }
\label{fig:bmf}
\end{figure}

\begin{figure}
\centering
\includegraphics[width=5in]{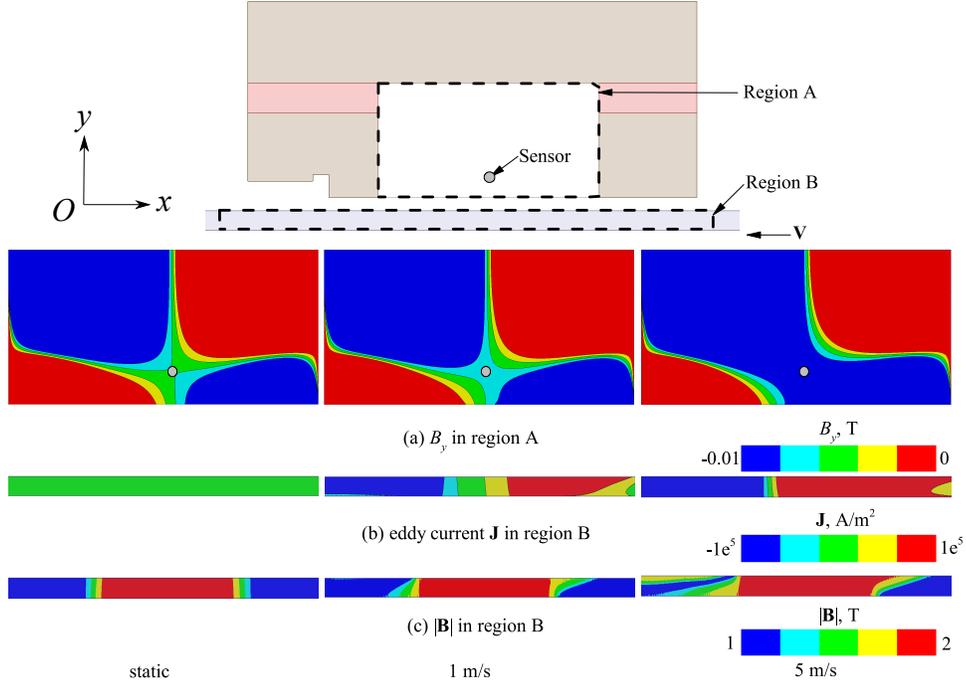}
%\vspace{-3mm}
\caption{(a) \textit{B}$_y$ distribution between two poles, (b) eddy current in the plate and (c) magnitude of \textbf{B} in the plate.} 
\label{fig:bmf_numerical_contour}
\end{figure}

Fig.\ref{fig:bmf} shows leakage signals with scanning location at different scanning velocities for the defect-free specimen. Results obtained from both experiments (a) and simulations (b).
Experimentally, as shown in Fig.\ref{fig:bmf} (a), the results indicated that leakage signals (axis component: \textit{B}$_y$), represented by the voltage, are approximately constant when the system is moving with a given velocity: e.g. after 1100 mm in the figure. Small fluctuations appear due to the specimen surface condition. Constant signals indicate that \textit{B}$_y$ retains a constant magnitude whilst scanning the defect free plate. Further, voltage magnitude increases as the scanning velocity is increased.\\
Two dimensional simulations were conducted to gain understanding into the mechanisms causing the results. Fig.\ref{fig:bmf} (b)  shows simulation results of \textit{B}$_y$ at different scanning velocities for a defect free specimen. It depicts that $|B_y|$ increases as the scanning velocity is increased, which is in an agreement with the experimental results. This phenomenon can be understood as follows: as the velocity is increased, the eddy current in the specimen is increased. The increased eddy current influences and distorts the magnetic field further, as shown in Fig.\ref{fig:bmf_numerical_contour}. 
As discussed in section \ref{sec:intro}, the magnetic Reynolds number is of the order of 100 and the secondary magnetic field, which is generated by the eddy current, distorts the original magnetic field. This distortion further increases the background magnetic field \textit{B}$_y$ in the vicinity of the sensor.
\subsection{Overview of leakage signals: without and with acceleration}
\label{sec:by}
\begin{figure}
\centering
\begin{subfigure}[b]{0.49\textwidth}
\includegraphics[width=\textwidth]{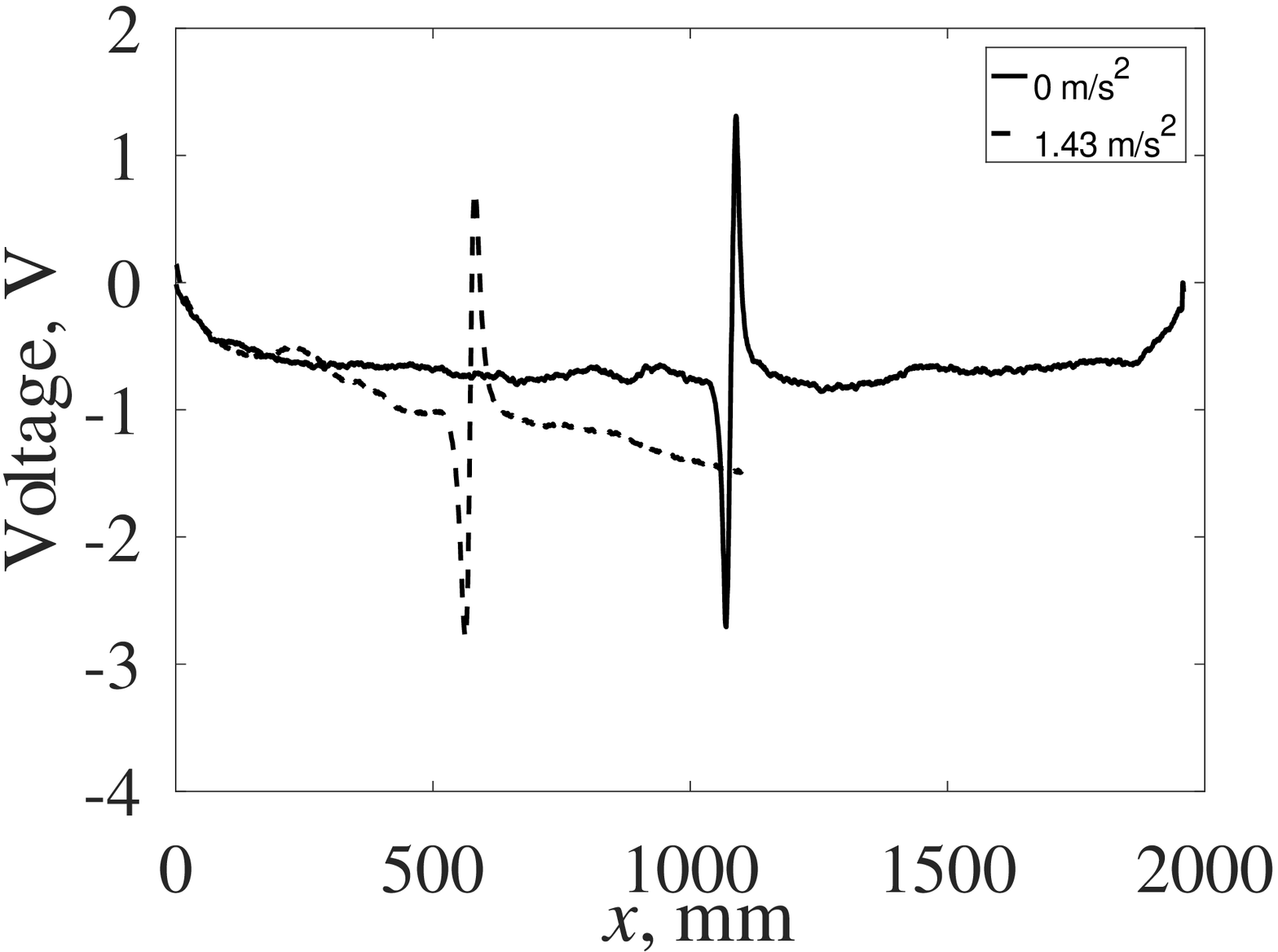}
\caption{4.8N}
\end{subfigure}
\begin{subfigure}[b]{0.49\textwidth}
\includegraphics[width=\textwidth]{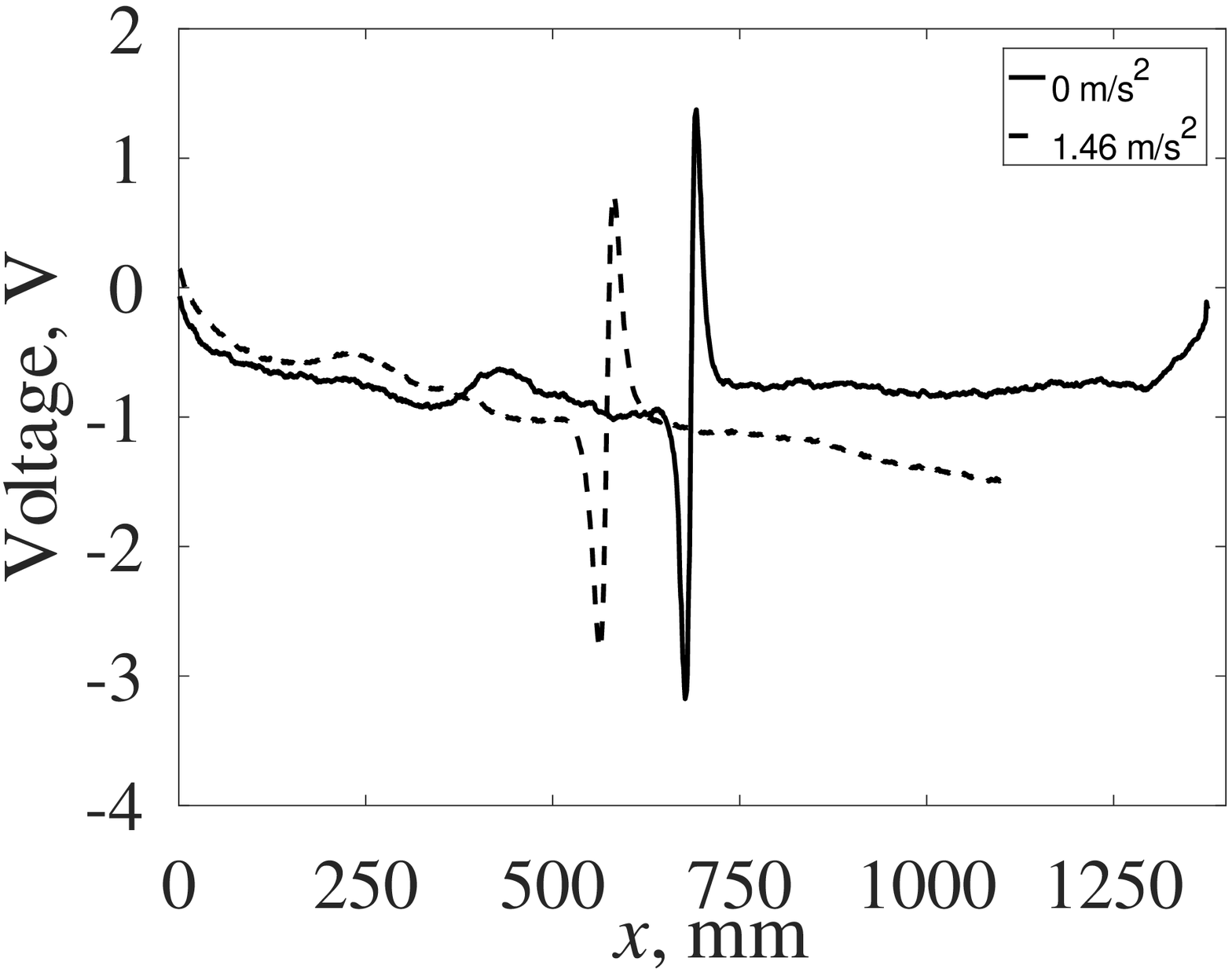}
\caption{4.8F}
\end{subfigure}
\begin{subfigure}[b]{0.49\textwidth}
\includegraphics[width=\textwidth]{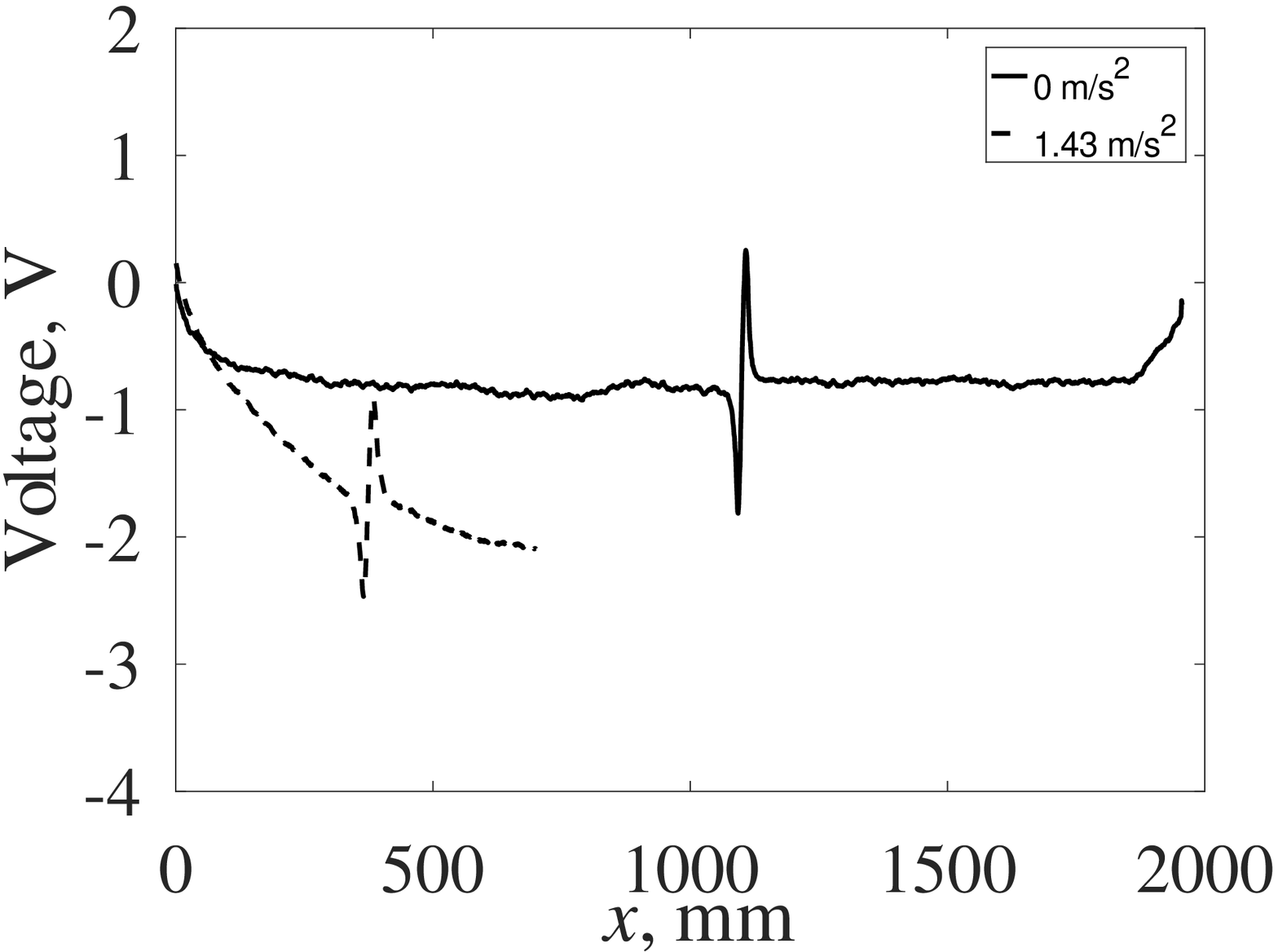}
\caption{2.4N}
\end{subfigure}
\begin{subfigure}[b]{0.49\textwidth}
\includegraphics[width=\textwidth]{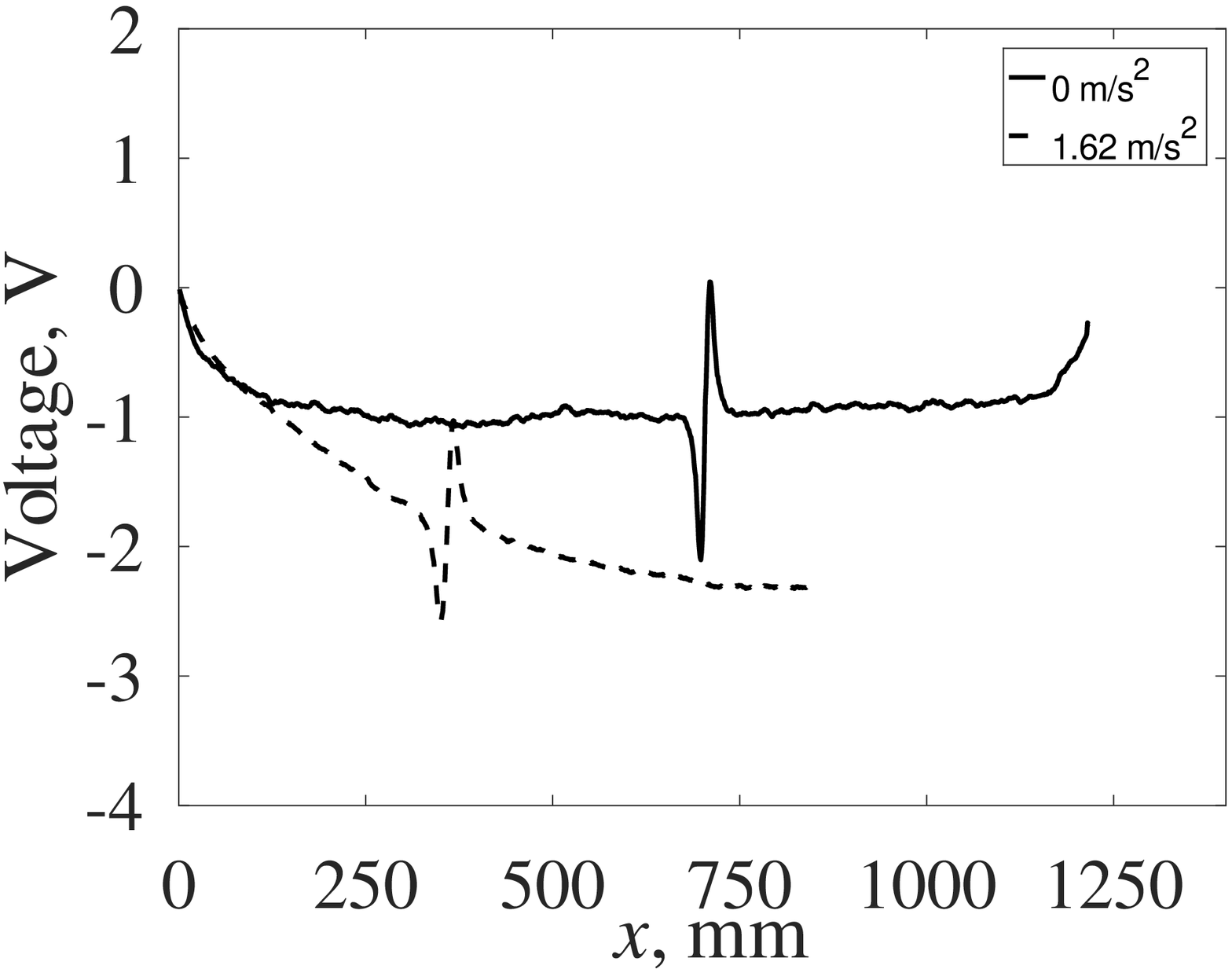}
\caption{2.4F}
\end{subfigure}
\caption{Experimental results: leakage signals for the cases both with and without scanning acceleration. The notation of 4.8N(or F) indicates \underline{N}ear (or \underline{F}ar) side defect with 4.8 mm defect depth.} 
\label{fig:overview}
\end{figure}

Fig.\ref{fig:overview} shows the overview of the leakage signal for the cases both with and without scanning system accelerations.
The results show that the scanning acceleration of the system does not change the trend of the leakage signal: peaks appear when the sensor meets defect edges. 
\subsection{Acceleration effect}
\label{sec:ae}
\subsubsection{P-p value comparison at similar scanning velocity}
\begin{figure}
\centering
\includegraphics[width=3.5in]{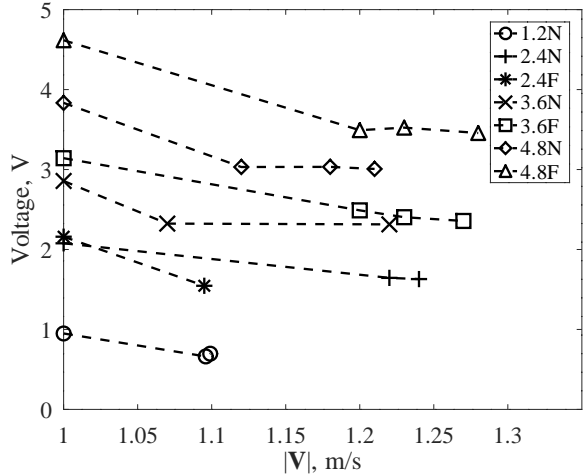}
%\vspace{-3mm}
\caption{Peak to peak value variations for different trials. The p-p value decreases when the system is accelerating. Cases with $\mid$\textbf{V}$\mid$= 1 m/s have a constant scanning velocity.}
\label{fig:aeffectsamev}
\end{figure}

Fig.\ref{fig:aeffectsamev} shows the peak to peak (p-p) value variations of \textit{B}$_y$ for trials both without acceleration (Trial B) and with acceleration (Trial C) at similar velocities for different maximum defect depths.
For both with and without acceleration trials, the p-p value increases as the defect depth is increased. It is also observed that defects in the bottom surface result in a higher magnitude of leakage compared to the same defect depth in the top surface. The  p-p values decreased for all results in Trial C compared to Trial B results with a similar velocity. The details of velocity and accelerations values for the points in Fig.\ref{fig:aeffectsamev} are listed in Tab.\ref{tab:samevdifferentable}
\begin{table}[htbp!]
\caption{Details of velocity and accelerations values for the points in Fig.\ref{fig:aeffectsamev}.}
\label{tab:samevdifferentable}
%	\vspace{-2mm} 
\begin{center}
\begin{tabular}{llllllll}
\hline
&1.2N&2.4N&2.4F&3.6N&3.6F&4.8N&4.8F\\
\hline					
$\mid$$\textbf{V}_0$$\mid$, m/s&1&1&1&1&1&1&1\\			
$\mid$$\textbf{a}_0$$\mid$, m/s$^2$&0&0&0&0&0&0&0\\			
\hline
$\mid$$\textbf{V}_1$$\mid$, m/s&1.09&1.22&1.10&1.07&1.20&1.12&1.20\\			
$\mid$$\textbf{a}_1$$\mid$, m/s$^2$&3.67&1.14&1.62&2.10&0.77&1.99&1.46\\		
\hline
$\mid$$\textbf{V}_2$$\mid$, m/s&1.10&1.24&-&1.22&1.23&1.18&1.23\\			
$\mid$$\textbf{a}_2$$\mid$, m/s$^2$&1.06&1.78&-&2.68&4.04&2.72&2.58\\			
\hline
$\mid$$\textbf{V}_3$$\mid$, m/s&-&-&-&-&1.27&1.21&1.28\\			
$\mid$$\textbf{a}_3$$\mid$, m/s$^2$&-&-&-&-&4.42&2.45&2.21\\			
\hline
\end{tabular}
\end{center}
\end{table}
This indicates that the presence of acceleration decreases the leakage signal. The magnitude of the p-p value reduction is in the range of 19\% to 28\%. These findings could help to optimise the MFL system to capture the defect feature precisely if defects are scanned during a phase of machine acceleration. It also shows the importance of conducting measurements at constant velocity, wherever possible.
\subsubsection{Acceleration magnitude influence}
\begin{figure}
\centering
\begin{subfigure}[b]{0.49\textwidth}
\includegraphics[width=\textwidth]{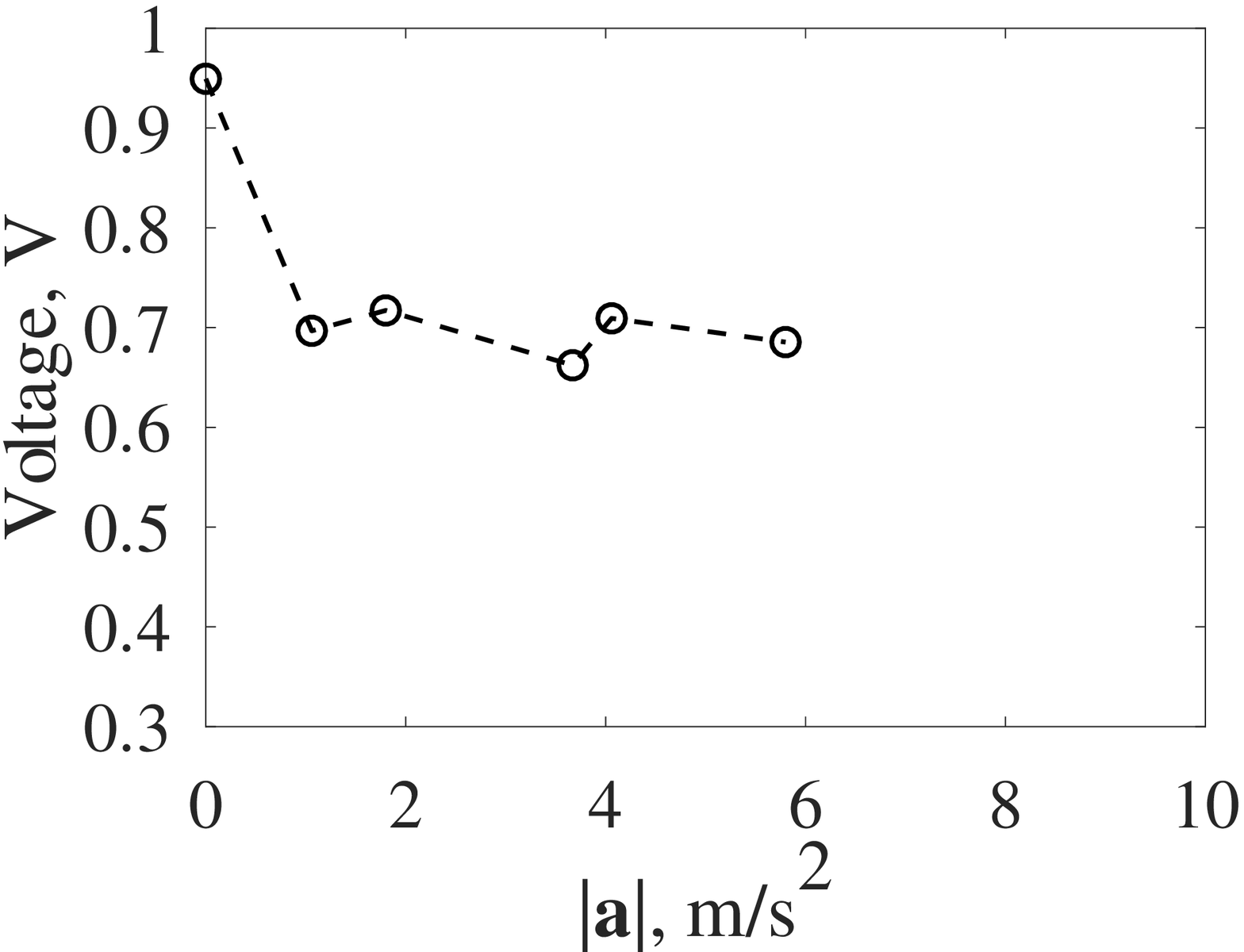}
\caption{1.2N}
\end{subfigure}
\begin{subfigure}[b]{0.49\textwidth}
\includegraphics[width=\textwidth]{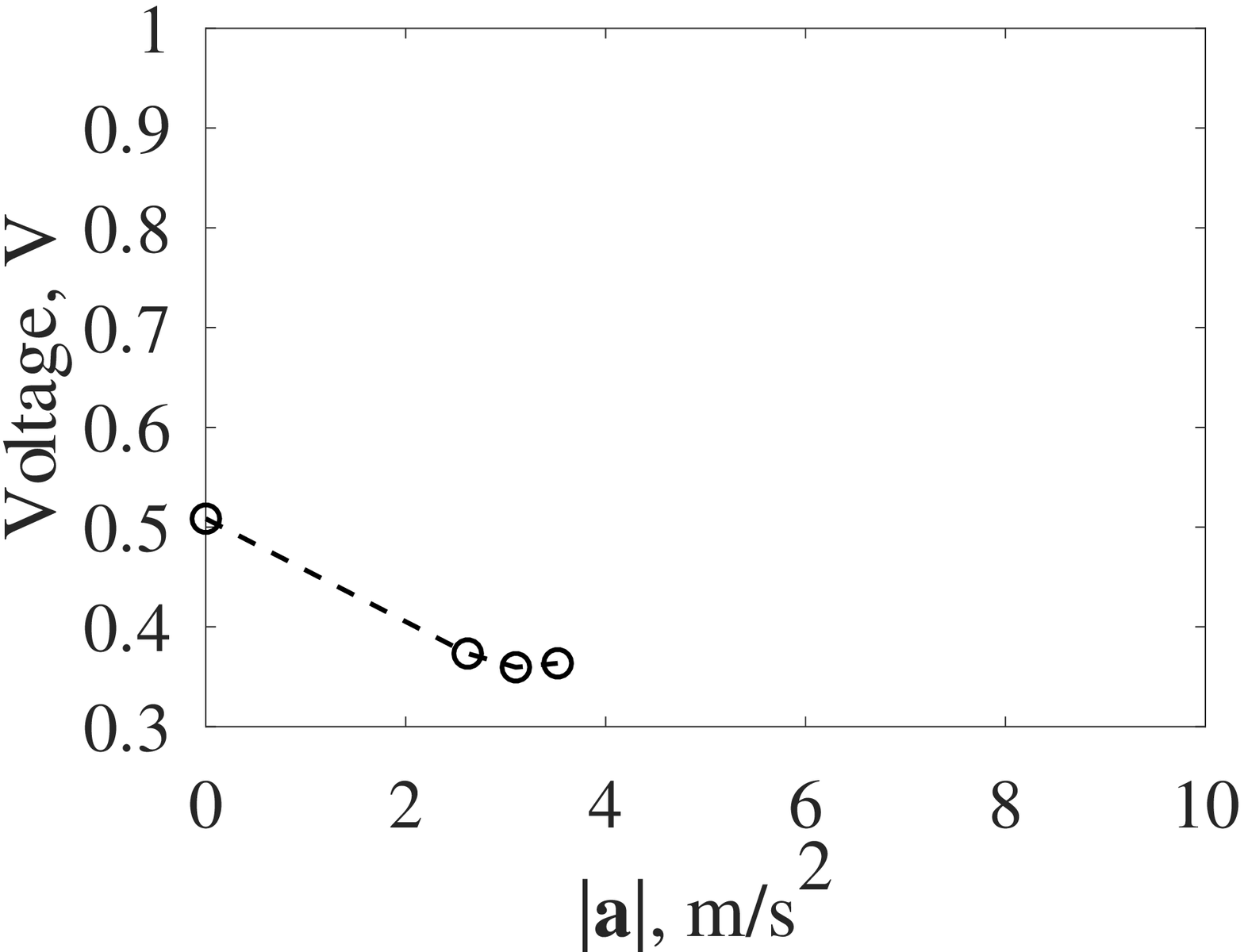}
\caption{1.2F}
\end{subfigure}
\begin{subfigure}[b]{0.49\textwidth}
\includegraphics[width=\textwidth]{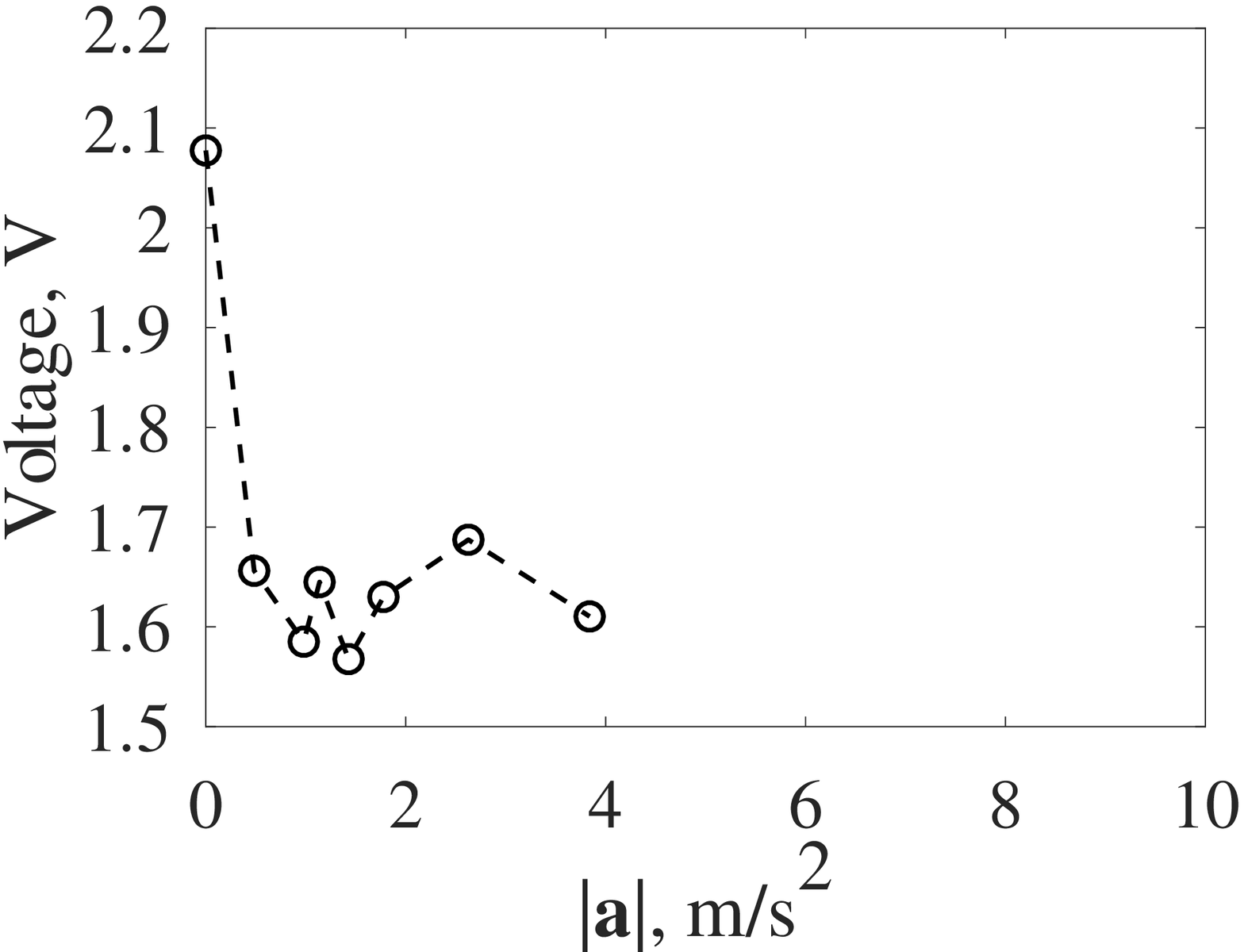}
\caption{2.4N}
\end{subfigure}
\begin{subfigure}[b]{0.49\textwidth}
\includegraphics[width=\textwidth]{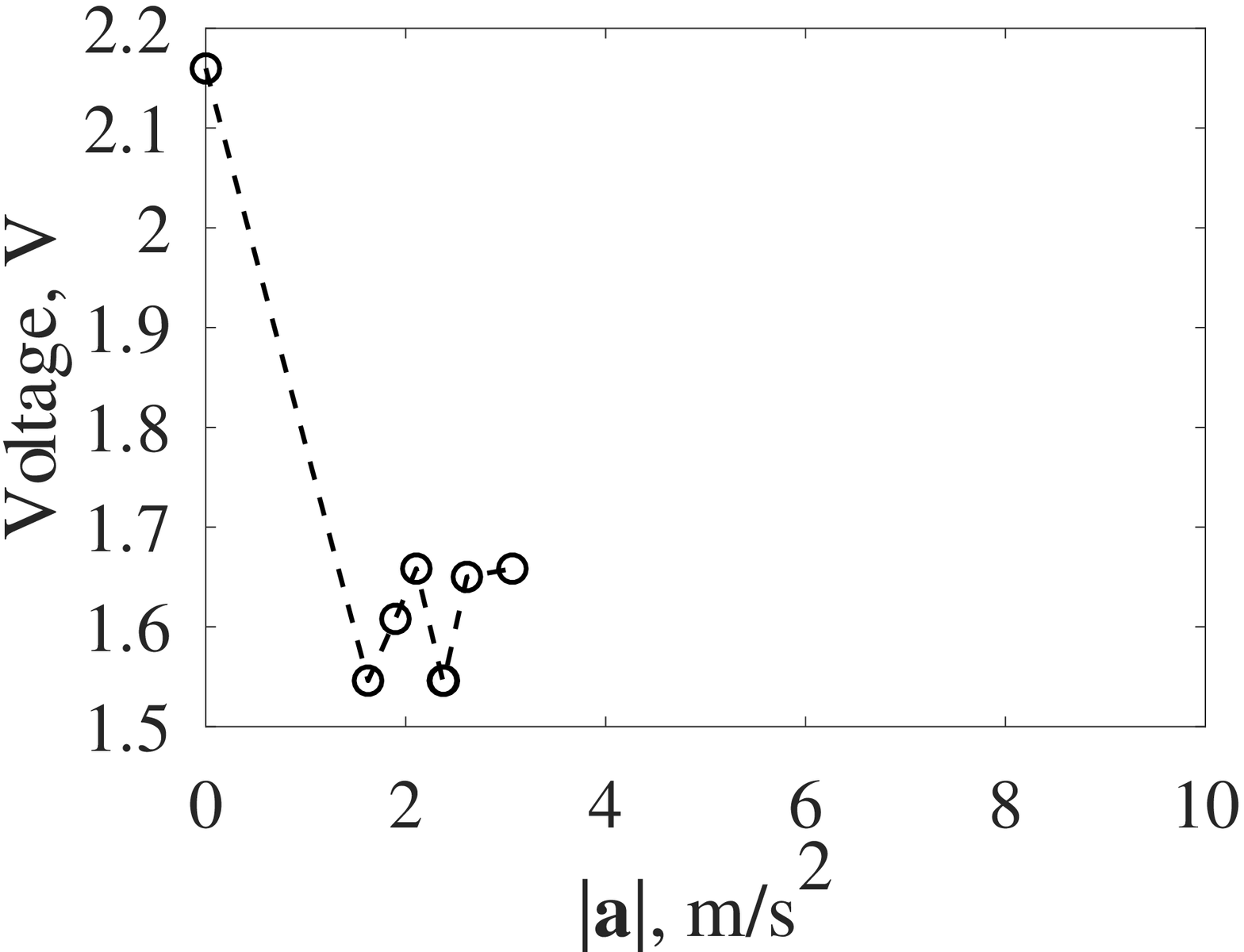}
\caption{2.4F}
\end{subfigure}
\begin{subfigure}[b]{0.49\textwidth}
\includegraphics[width=\textwidth]{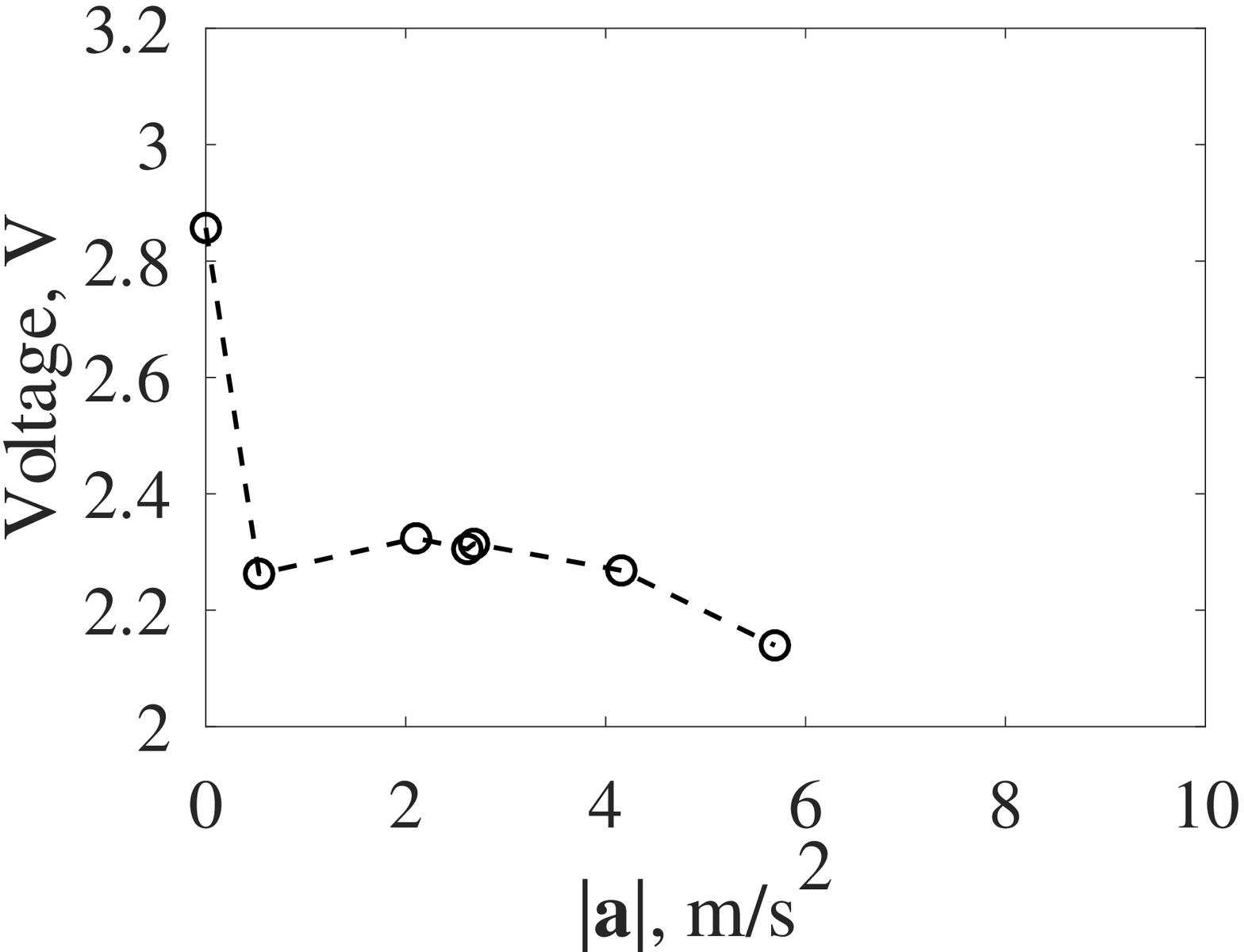}
\caption{3.6N}
\end{subfigure}
\begin{subfigure}[b]{0.49\textwidth}
\includegraphics[width=\textwidth]{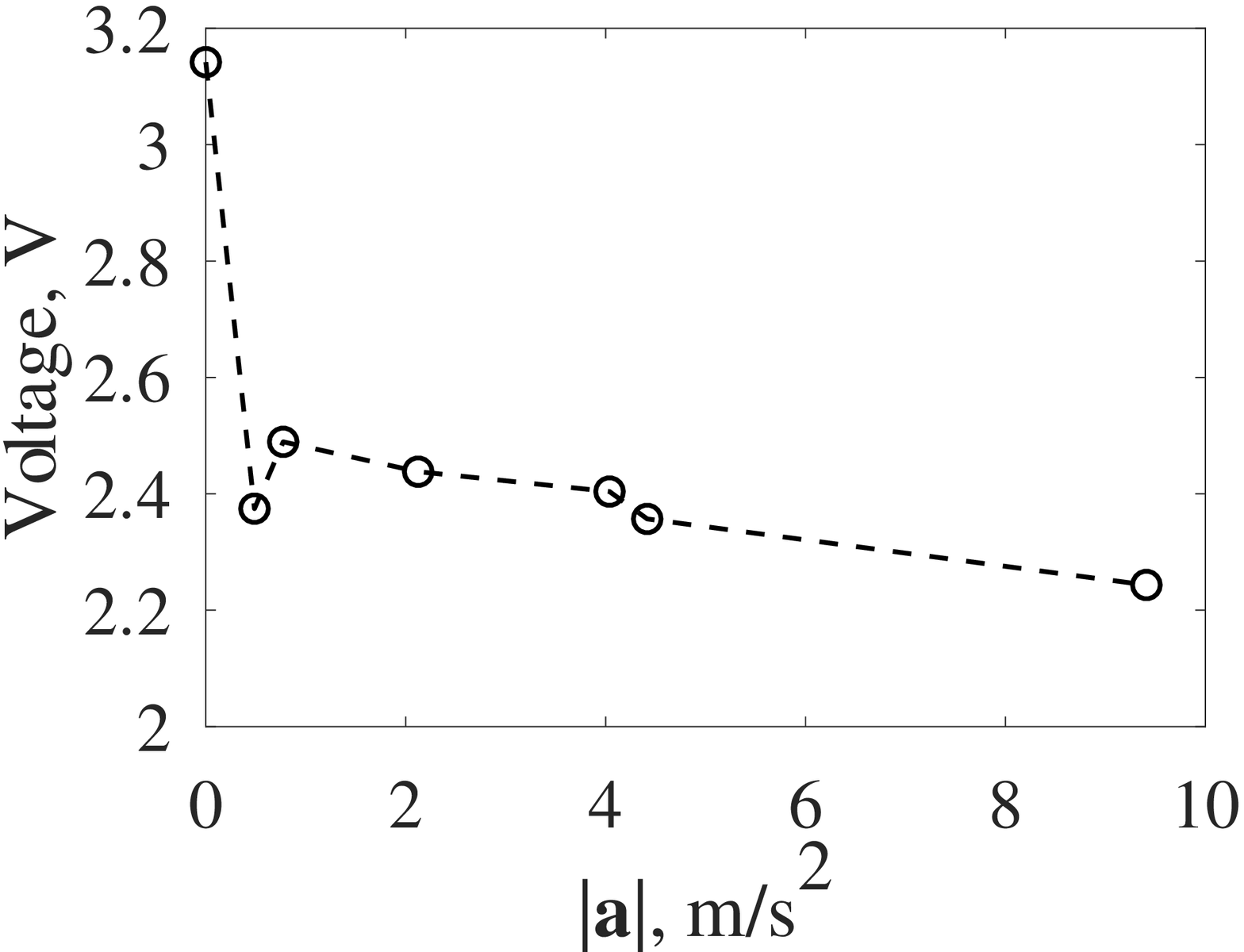}
\caption{3.6F}
\end{subfigure}
\begin{subfigure}[b]{0.49\textwidth}
\includegraphics[width=\textwidth]{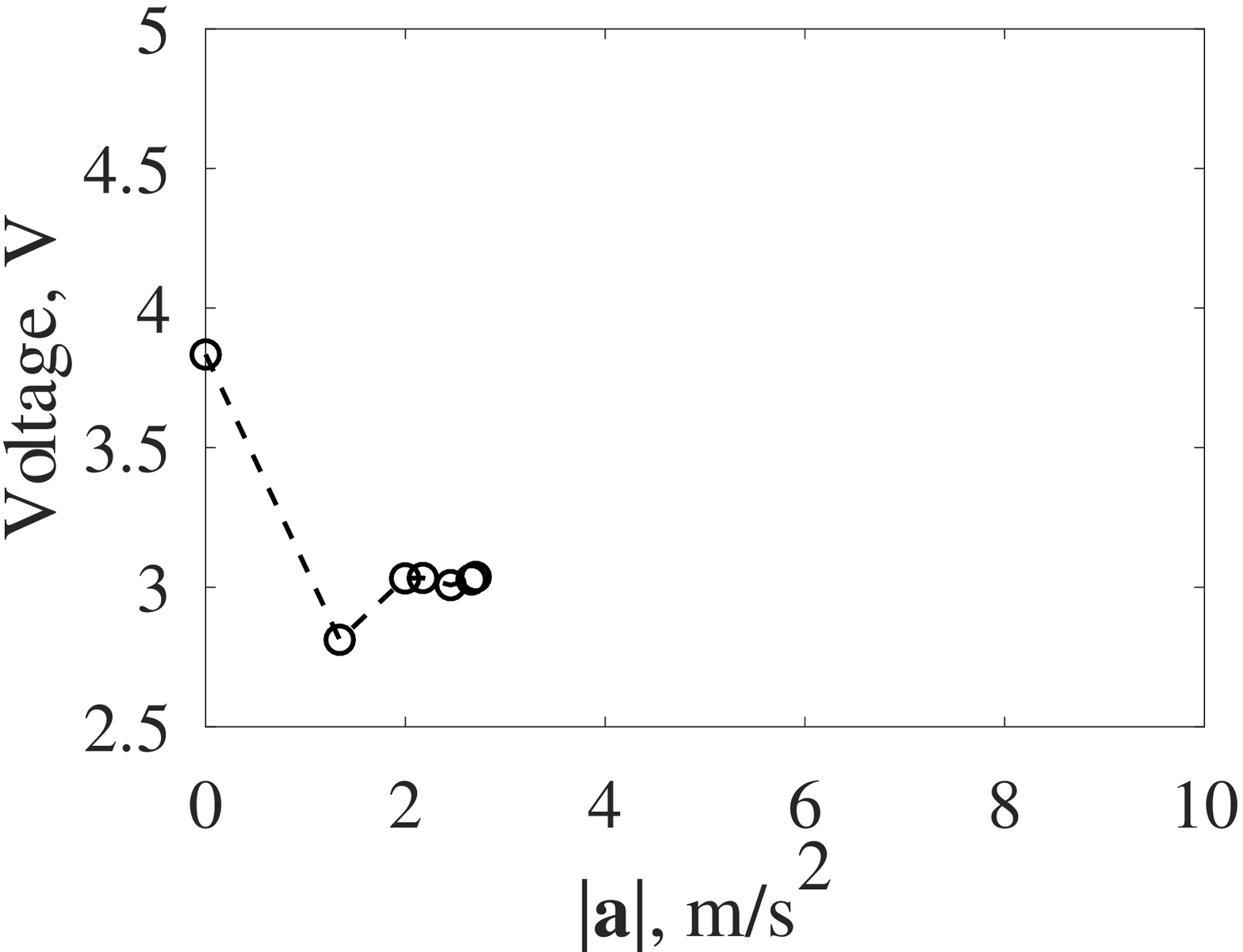}
\caption{4.8N}
\end{subfigure}
\begin{subfigure}[b]{0.49\textwidth}
\includegraphics[width=\textwidth]{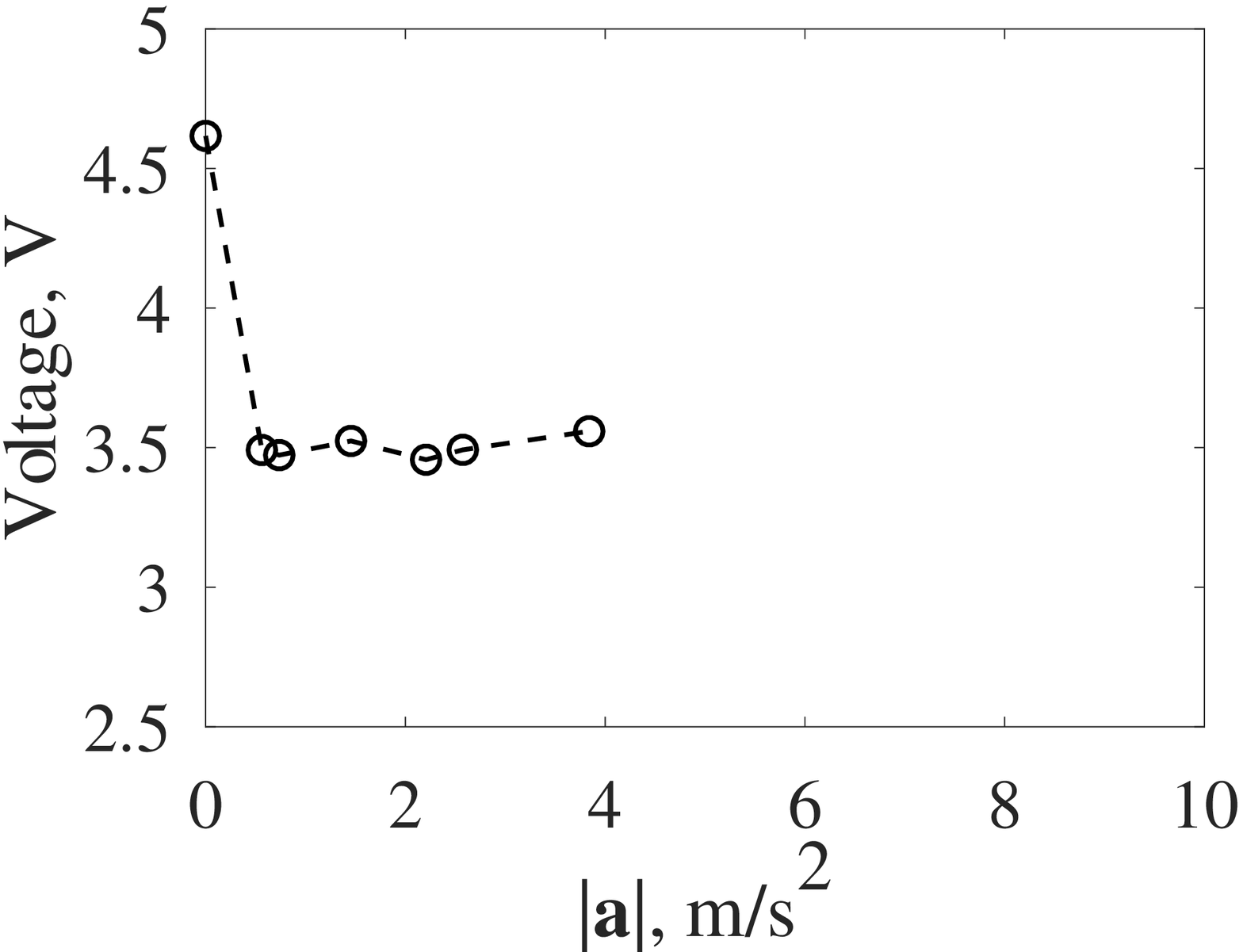}
\caption{4.8F}
\end{subfigure}
\caption{Experimental results: Peak to peak value variation for different trials.} 
\label{fig:aeffectdifferenta}
\end{figure}

Fig.\ref{fig:aeffectdifferenta} shows the p-p value variations with different acceleration values.
The results from all the experimental trials shows that the magnitude of the reduction of the leakage signal is not sensitive to the magnitude of the acceleration. 
\subsubsection{Initial discussion}
\begin{figure}
\centering
\begin{subfigure}[b]{0.48\textwidth}
\includegraphics[width=\textwidth]{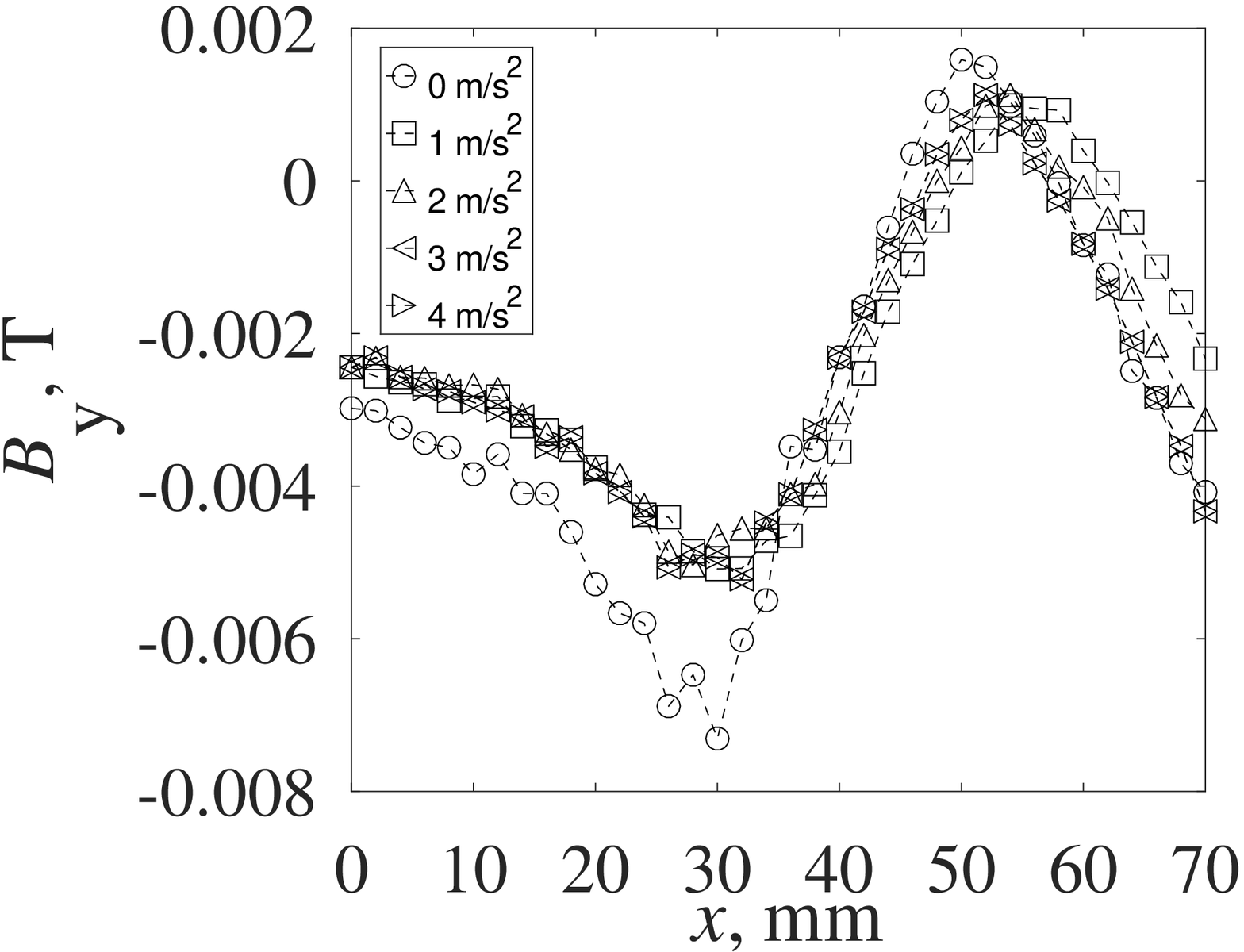}
\caption{}
\end{subfigure}
\begin{subfigure}[b]{0.5\textwidth}
\includegraphics[width=\textwidth]{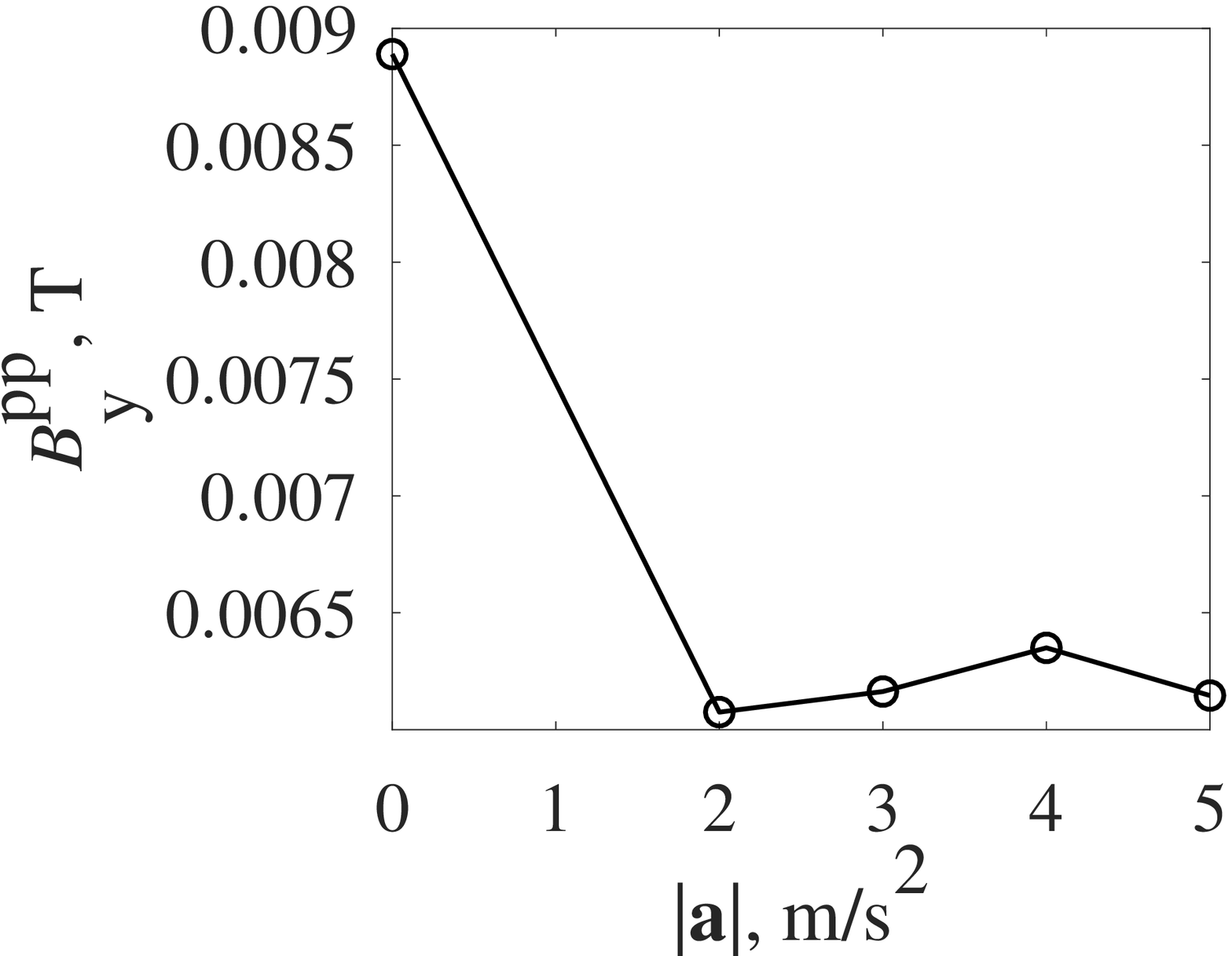}
\caption{}
\end{subfigure}
\caption{Simulation results: \textit{B}$_y$ signal (a) and peak to peak value variations with \textbf{a} (b). The peak to peak value drops when acceleration is present.}
\label{fig:aeffectsimu}
\end{figure}

\begin{figure}
\centering
\includegraphics[width=5in]{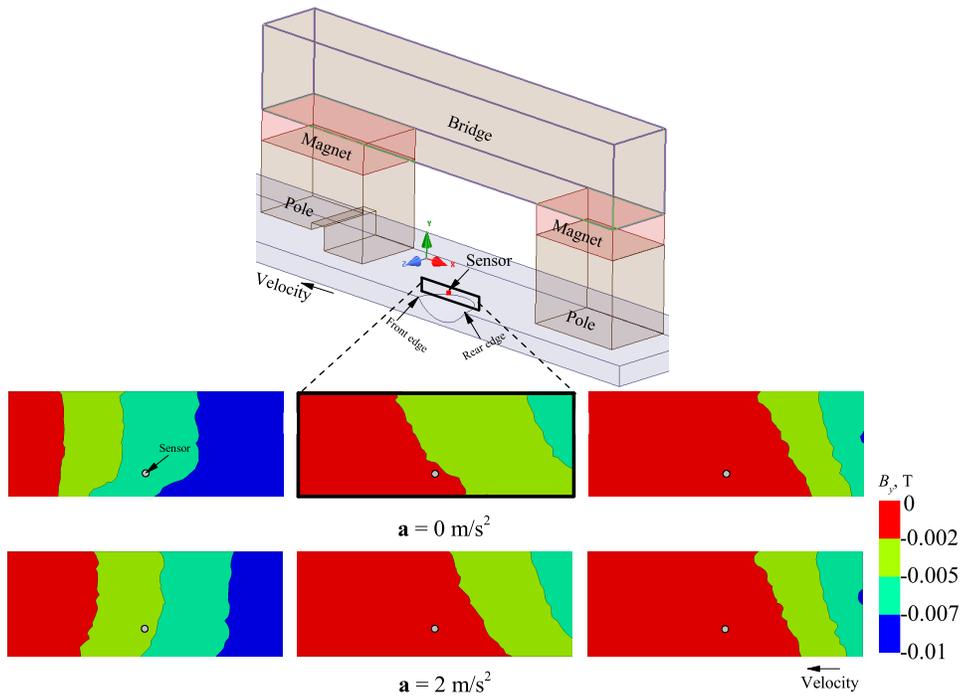}
%\vspace{-3mm}
\caption{\textit{B}$_y$ contours in the vicinity of sensor. Contours plotted when the sensor meets the front defect edge ($Left$), the defect middle ($Middle$) and the rear defect edge ($right$). Different magnetic field distortions are presented for the cases with and without system scanning acceleration. The acceleration, \textbf{a}, has the same direction with velocity.}
\label{fig:aeffectsimu3d}
\end{figure}

The experimental results show two main effects in terms of the acceleration influence:
\begin{enumerate}
\item The presence of the acceleration reduces the leakage signal by around 20\% to 30\%; 
\item The magnitude of the reduction in leakage signal is not sensitive to the magnitude of the acceleration (up to  6 m/s$^2$).
\end{enumerate}
The results also indicated that the defect features cannot be captured precisely if the defect is scanned whilst the MFL system is accelerating.\\
To help understand the phenomenon, preliminary 3D simulations were conducted. Fig.\ref{fig:aeffectsimu} shows the \textit{B}$_y$ distribution with scanning time (left) and the p-p value for different system accelerations (right). The 3D numerical results showed that the presence of the acceleration does not change the trend of \textit{B}$_y$ signal, as shown in  Fig.\ref{fig:aeffectsimu} ($left$). 
In terms of the leakage value, Fig.\ref{fig:aeffectsimu} ($right$), a clear reduction in magnitude is seen for an accelerating system. The simulation results also show that the magnitude of leakage reduction does not change when the system acceleration is  further increased. These results are in agreement with the experimental results.\\
Fig.\ref{fig:aeffectsimu3d} shows \textit{B}$_y$ distribution for the case with and without acceleration at the moment when the sensor meets the defect front edge, the defect middle and the defect rear edge, respectively.
The results indicate that higher magnitude of \textit{B}$_y$ is obtained for the case without acceleration for all three positions, especially for the position when the sensor meets the front edge of the defect. This is in agreement with the results shown in Fig.\ref{fig:aeffectsimu} ($left$). These \textit{B}$_y$ differences result in the difference of p-p values.
\section{Conclusions and future work}
\label{sec:con}
This work investigated the influence of magnetic flux leakage system scanning acceleration on the leakage signal. The work was conducted experimentally and numerically. The main findings are summarised as follows:
\begin{enumerate}
\item The magnitude of background magnetic field ($y$-axis component: direction perpendicular to scanning velocity) increases with increased scanning velocity. We conjecture this is due to the eddy current effect, which is generated in the specimen.
\item An accelerating magnetic leakage system does not change the general features of \textit{B}$_y$ distribution. 
\item Both the experimental and the numerical findings showed that the leakage signal (under acceleration), which is evaluated by using peak to peak value, will reduce in a range of 20\% to 30\%. Preliminary simulation results showed that the largest difference is present when the sensor meets the front edge of the defect for cases with and without acceleration.
\item The results indicate that magnetic flux leakage system should ideally account for signal measurements during the start up and stopping stages, and where possible measurements should be obtained when the system is moving with constant velocity.
\end{enumerate} 
Future work will focus on a further understanding of the acceleration effect by using 3D simulations in greater depth.
\section*{Acknowledgements}
The authors would like to acknowledge the ASTUTE 2020 (Advanced Sustainable Manufacturing Technologies) operation supporting manufacturing companies across Wales, which has been part-funded by the European Regional Development Fund through the Welsh Government and the participating Higher Education Institutions. The authors would also like to acknowledge the contribution of Eddyfi Technologies.

\end{document}